# Diffusion and equilibration of site-preferences following transmutation of tracer atoms

Gary S. Collins

Dept of Physics and Astronomy, Washington State University, Pullman, WA, 99164, USA

collins@wsu.edu

**Abstract.** Using the method of perturbed angular correlation of gamma rays, diffusional jump-frequencies of probe atoms can be measured through relaxation of the nuclear quadrupole interaction. This was first shown in 2004 for jumps of tracer atoms that lead to reorientation of the local electric field-gradient, such as jumps on the connected $\alpha$-sublattice in the $L1_2$ crystal structure. Studies on many such phases using the $^{111}$In/Cd PAC probe are reviewed in this paper. A major finding from a 2009 study of indides of rare-earth elements, $In_3R$, was the apparent observation of two diffusional regimes: one dominant for heavy-lanthanide phases, R= Lu, Tm, Er, Dy, Tb, Gd, that was consistent with a simple model of vacancy diffusion on the In $\alpha$-sublattice, and another for light-lanthanides, R= La, Ce, Pr, Nd, that had no obvious explanation but for which several alternative diffusion mechanisms were suggested. It is herein proposed that the latter regime arises not from a diffusion mechanism but from transfer of Cd-probes from In-sites where they originate to R-sites as a consequence of a change in site-preference of $^{111}$Cd-daughter atoms from In-sites to R-sites following transmutation of $^{111}$In. Support for this transfer mechanism comes from a study of site-preferences and jump-frequencies of $^{111}$In/Cd probes in $Pd_3R$ phases. Possible mechanisms for transfer are described, with the most likely mechanism identified as one in which Cd-probes on $\alpha$-sites transfer to interstitial sites, diffuse interstitially, and then react with vacancies on $\beta$-sites. Implications of this proposal are discussed. For indides of heavy-lanthanide elements, the Cd-tracer remains on the In-sublattice and relaxation gives the diffusional jump-frequency.

## Introduction

There is considerable interest in diffusion of atoms in solids [1]. Fluctuating electric field-gradients at nuclei of diffusing hyperfine probe atoms cause relaxation of the nuclear quadrupole interaction. Given an appropriate microscopic model for a diffusion mechanism, the relaxation can be analyzed to yield precise values of the mean jump-frequency of the probe nuclide [2]. Using the method of perturbed angular correlation of gamma rays (PAC), this was first observed in equilibrium studies of $^{111}$In/Cd probes in the $L1_2$ phase $In_3La$ [2]. $^{111}$In decays by electron-capture to the second excited state of $^{111}$Cd that, in turn, promptly decays to the 247 keV PAC level of $^{111}$Cd and then to the ground state. The PAC level has a mean lifetime of 120 ns. An electric field-gradient at the nuclear site exerts, in effect, a torque on the quadrupole moment of the nucleus that leads to a time-dependent perturbation during the lifetime of the PAC level. For a detailed description of PAC spectroscopy and methods, see [3] and papers cited below.

**Nuclear relaxation caused by diffusion.** Fig. 1 shows the $L1_2$ crystal structure for a generic $A_3B$ phase, with A-atoms at face-centers forming a sublattice with near-neighbor connections known as the $\alpha$-sublattice. The arrows illustrate two jumps of atoms caused by rapid passage of vacancies on the $\alpha$-sublattice. This simple diffusion mechanism is known as the $\alpha$-sublattice vacancy diffusion mechanism [22, 4]. It is energetically favored since no new point defects are created in a jump;

there is a simple exchange in positions of the jumping atom and vacancy, as in pure metals. In none of the present studies have signals been observed that could be attributed to the EFG disturbance of a nearby vacancy. It is therefore assumed that diffusional vacancies make rapid transits through neighborhoods of the probe atoms, only shuffling their positions. As a consequence, EFG changes arise solely from the difference between the local atomic arrangements before and after the jump. In the course of each atomic jump on the sublattice, the local tetragonal axis of symmetry (*x*, *y* or *z* in Fig. 1) reorients by 90$^o$. The principal axis of the electric-field gradient (EFG) lies along the local tetragonal axis and, in a manner of speaking, the quadrupole moment of the probe's nucleus precesses about that axis.

Fig. 2 exhibits PAC spectra measured for $^{111}$In/Cd probes in the L1$_2$ phase In$_3$La at the indicated temperatures [2]. When the residence time of the probe atom at a site is less than the lifetime of the PAC level, jumps lead to decoherence in the quadrupolar precessions that appears as "damping" of the time-domain perturbation. With increasing temperature, spectra are seen to evolve from the nearly static quadrupolar perturbation function at 156$^o$C, through a damped function at 261$^o$C, to a function exhibiting maximum relaxation at ~340$^o$C, and eventually to a motionally-averaged perturbation at 629$^o$C that exhibits no periodicity for a very high jump-frequency. In the slow-fluctuation regime, at temperatures below 340°C, the observed spectrum appears in good approximation as the product of a static perturbation function $G_{static}(t)$ and an exponential relaxation,

$$G_2(t) \cong \exp(-\lambda t) G_{static}(t), \quad (1)$$

in which $\lambda$ is a relaxation frequency caused by fluctuating EFGs. All spectra such as in Fig. 2 were fitted using exact numerical simulations of fluctuating EFGs arising from probe atoms making jumps on the In-sublattice via the $\alpha$-sublattice diffusion mechanism, leading to precise values of the jump frequency for the assumed mechanism [7]. Agreement between the theoretical lineshapes and spectral data in Fig. 2 is excellent, indicating that all probes experience the same relaxation and that, for example, there are not two ensembles of probes experiencing different relaxations. In the slow fluctuation regime, for 90° reorientations of the EFG, the relaxation frequency is in good approximation equal to the mean jump-frequency $w$ of the probe on the $\alpha$-sublattice, $\lambda \cong w$, in which $w$ is the inverse of the mean residence-time of the probe on a site [5]. Such a relationship cannot be assumed for another diffusion mechanism; for example, $\lambda < w$ if the EFG reorients through an angle less than 90° in a jump.

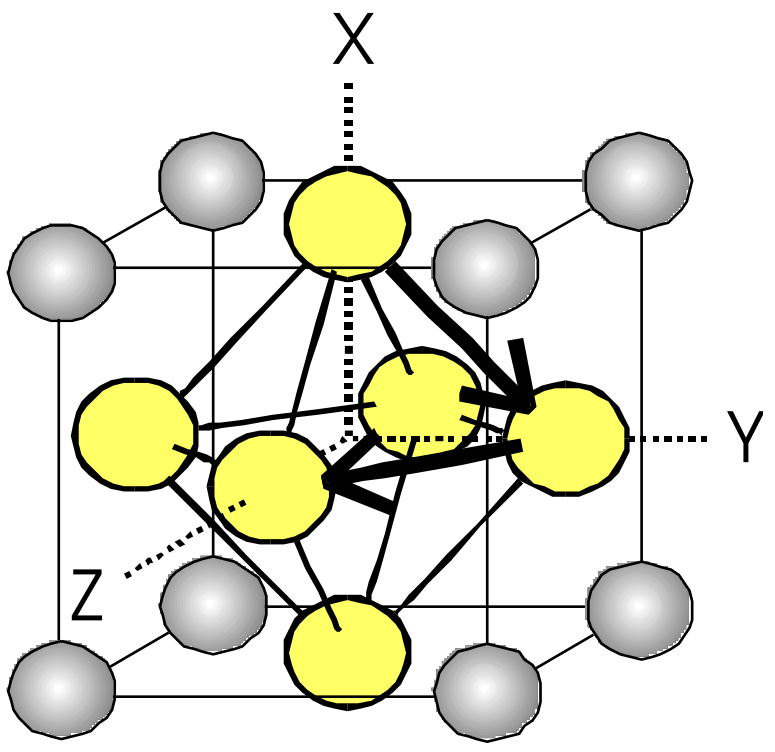

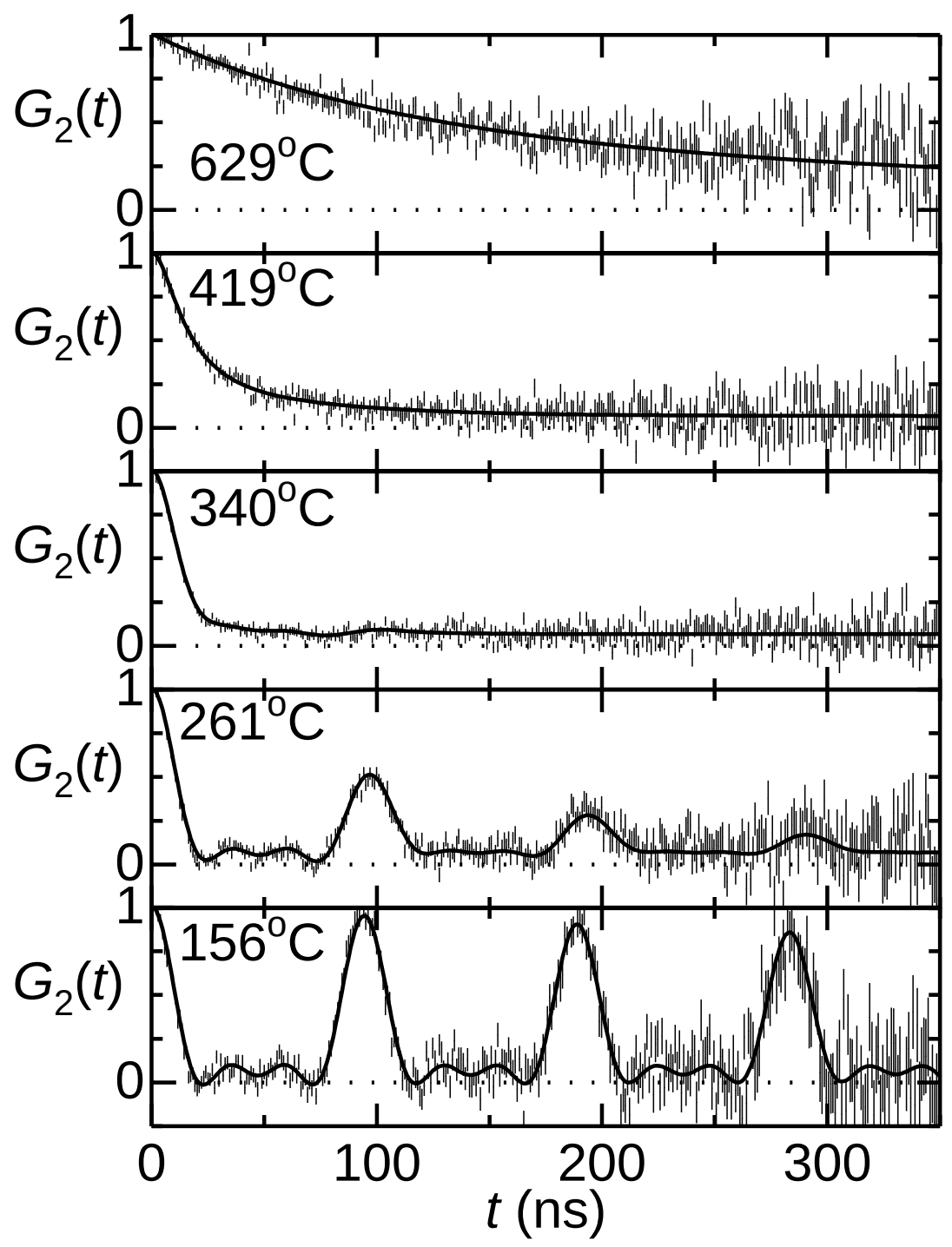

Fig. 1. (left) L1$_2$ crystal structure. The sublattice of face-centered sites has principal axes for the electric field-gradient tensor (EFG) at each site pointing along local tetragonal X, Y or Z directions. Diffusion of vacancies on that sublattice leads to jumps of a probe atom, shown by the arrows, with reorientation of the EFG tensor by 90$^o$ in each jump. The reorientations of the EFG cause decoherence, or relaxation, of the nuclear quadrupole interaction.

Fig. 2. (right) PAC spectra for $^{111}$In/Cd probes in the L1$_2$ phase In$_3$La (from [2]). With increasing temperature, spectra are seen to evolve from a nearly static quadrupole perturbation function at 156$^o$C, through a function having maximum relaxation at 340$^o$C, to a motionally-averaged perturbation at 629$^o$C that exhibits no periodicity. These spectra were fitted using numerical simulations of stochastically fluctuating EFGs for probe atoms making jumps on the In-sublattice.

**Composition dependence of jump frequencies.** $In_3R$ phases appear in binary phase diagrams as line compounds at the stoichiometric 3:1 composition. From a practical point of view, this means that the width of the phase field is less than about 1 at%. To our knowledge, no precise measurements have been made of widths and bounding compositions of $L1_2$ phase fields for $In_3La$ or other $In_3R$ phases (R= rare-earth). PAC experiments to measure jump-frequencies of probe atoms were originally made in this laboratory using "stoichiometric" and presumably single-phase samples produced by arc-melting high-purity metal foils under argon in a small arc-furnace. However, repeated measurements exhibited poor reproducibility of jump frequencies [6]. This was eventually attributed to gradual changes in composition of ~100 mg spherical samples due to evaporation of indium from the binary alloy during day-long measurements at high temperature under high vacuum. Within a single-phase field, a shift in average composition leads to changes in mole fractions of intrinsic point defects, including vacancies on the two sublattices that are the direct mediators of diffusion. To avoid such drifts in composition, later experiments were carried out using pairs of samples prepared with average compositions leading to a large volume fraction of the $L1_2$ phase of interest as well as minor fractions of adjacent equilibrium phases. Typical mean sample compositions were 24 and 26 at.% of the rare-earth element. The adjacent phase may lead to a small site fraction of a second quadrupole interaction signal, but this never posed problems in analyzing signals from the $L1_2$ phase. Indeed, terminal indium metal is the more In-rich phase adjacent to $L1_2$ phases in all In-R systems. Indium melts at 157°C, and probe atoms in molten In exhibit only a motionally-averaged zero-frequency quadrupolar perturbation with no structure. By using samples containing a small fraction of a neighboring secondary phase, the composition of the $L1_2$ phase of interest was anchored at its boundary composition, even in the presence of modest evaporation of one alloy component, and reproducibility of jump-frequencies was much improved [6]. Finally, PAC involves measuring time-delayed coincidence spectra, so that source activity must be kept low to reduce accidental coincidences. $^{111}$In activity is available commercially in a carrier-free state, so that the initial mole fraction of $^{111}$In in a typical sample is only ~$10^{-11}$, much lower than mole fractions of structural or thermally activated intrinsic point defects.

Fig. 3 shows Arrhenius plots of jump frequencies $w$ of the In/Cd probes in $In_3La$, obtained from fits of nuclear relaxation assuming stochastic jumps via the $\alpha$-sublattice diffusion mechanism, with $w \cong \lambda$. The upper and lower data sets are from measurements on the samples having, respectively, In-rich and In-poor boundary compositions of the $L1_2$ phase. The composition dependence of the jump frequency gives insight into the diffusion mechanism. For the $\alpha$-sublattice diffusion mechanism, the jump-rate should be proportional to the mole-fraction of vacancies on the $\alpha$-sublattice. Mole-fractions of vacancy and antisite point-defects vary monotonically with the composition of a phase. In particular, the mole-fraction of In-vacancies will be greater in an In-poor sample than in an In-rich one, leading to the expectation of a greater jump-frequency at a given temperature in the In-poor sample than in the In-rich sample for the $\alpha$-sublattice vacancy diffusion mechanism. As can be seen from the figure, however, relaxation frequencies were much greater in In-rich samples. Therefore, the simple $\alpha$-sublattice diffusion mechanism can be ruled out by the observed dependence on composition [2, 7]. Three more complex diffusion mechanisms were discussed that are consistent with the composition dependence [2, 7]: a La-vacancy six-jump cycle, a divacancy mechanism in which bound In and rare-earth vacancy pairs make correlated jumps, and a conversion mechanism in which a host In atom, $In_\alpha$, reacts with a La-vacancy, $V_\beta$, to form a bound complex $V_\beta + In_\alpha \rightarrow V_\alpha : In_\beta$, with the vacancy jumping freely around the antisite atom and occasionally reconstituting the $V_\beta$. Activation enthalpies obtained from fits of the relaxation-frequency data in Fig. 3 are 0.53(1) eV and 0.81(1) eV for the In-rich and In-poor compositions, respectively, much smaller than what is typically observed for diffusion of tracers in intermetallic compounds [8]. Indeed, extrapolation of the fitted data for the In-rich sample (upper data set) to room-temperature suggests an extraordinarily high jump-frequency of 1 kHz [2]. The three mechanisms listed above seem implausible in light of such low activation enthalpies.

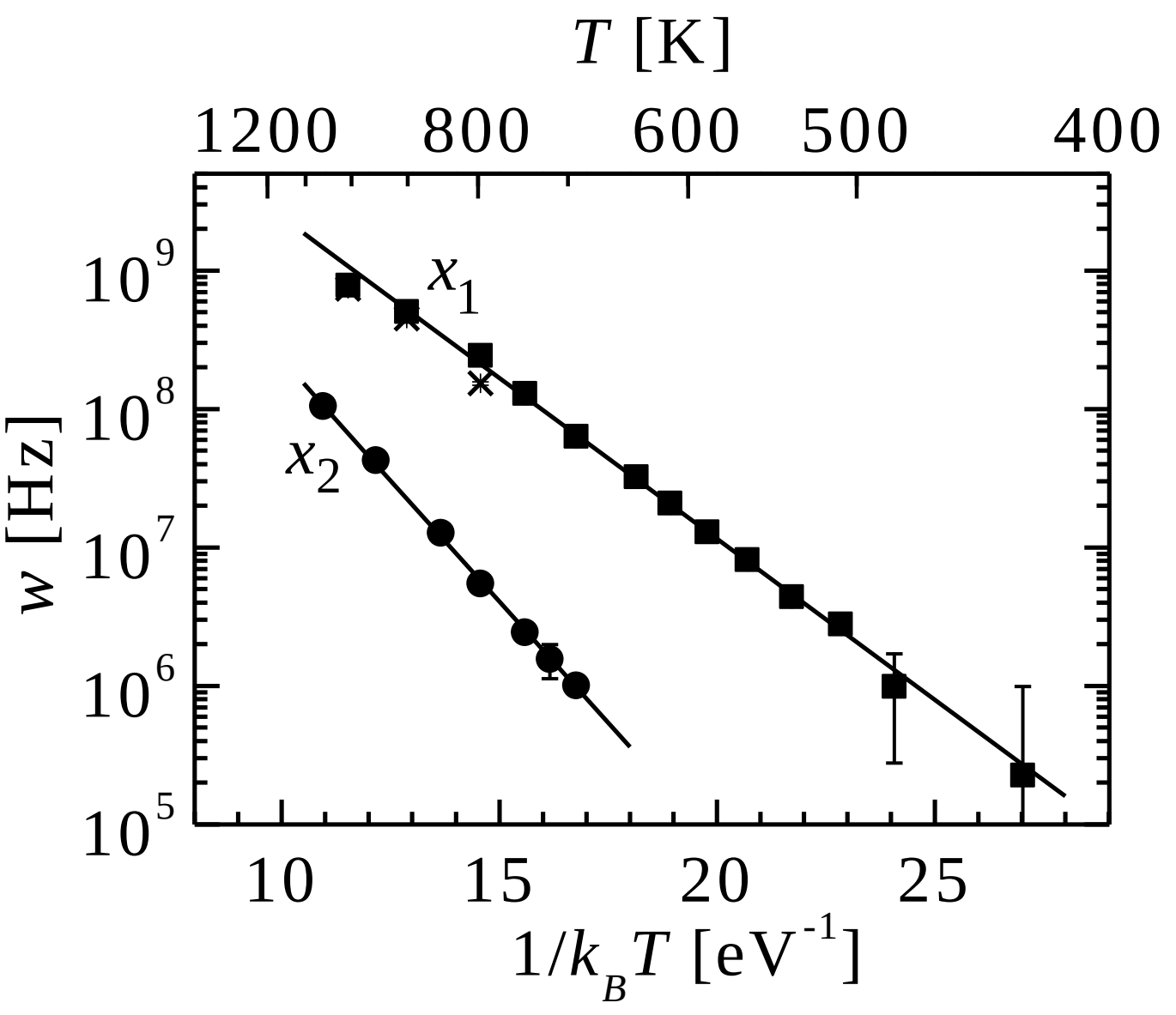

Fig. 3. Arrhenius plots of diffusional jump-frequencies of $^{111}$In/Cd tracer atoms in the $L1_2$ phase $In_3La$ attributed to probe atoms jumping on the In-sublattice through passage of rapidly diffusing vacancies (from [2]). Data sets are for measurements at opposing boundary compositions of the $In_3La$ phase: the In-richer ($x_1$) and In-poorer ($x_2$) boundary compositions, with fitted activation enthalpies 0.53(1) and 0.81(1) eV, respectively.

**Nuclear relaxation for In/Cd probes in the series of rare-earth indides.** Following the original experiments on $In_3La$, comprehensive measurements were made on all $In_3R$ rare-earth (R) phases for both In-rich and In-poor boundary compositions [7]. Results for 25 samples are summarized in Fig. 4, with each data point representing results of measurements on a sample at a series of temperatures. Values of jump-frequency were obtained by fitting the experimental nuclear relaxation assuming the $\alpha$-sublattice diffusion mechanism ($w \cong \lambda$). Fits of Arrhenius plots of jump-frequency $w$ such as shown in Fig. 3 yielded activation enthalpies $Q$ and jump-frequency prefactors $w_0$ according to

$$w = w_0 \exp(-Q / k_B T)\,. \qquad (2)$$

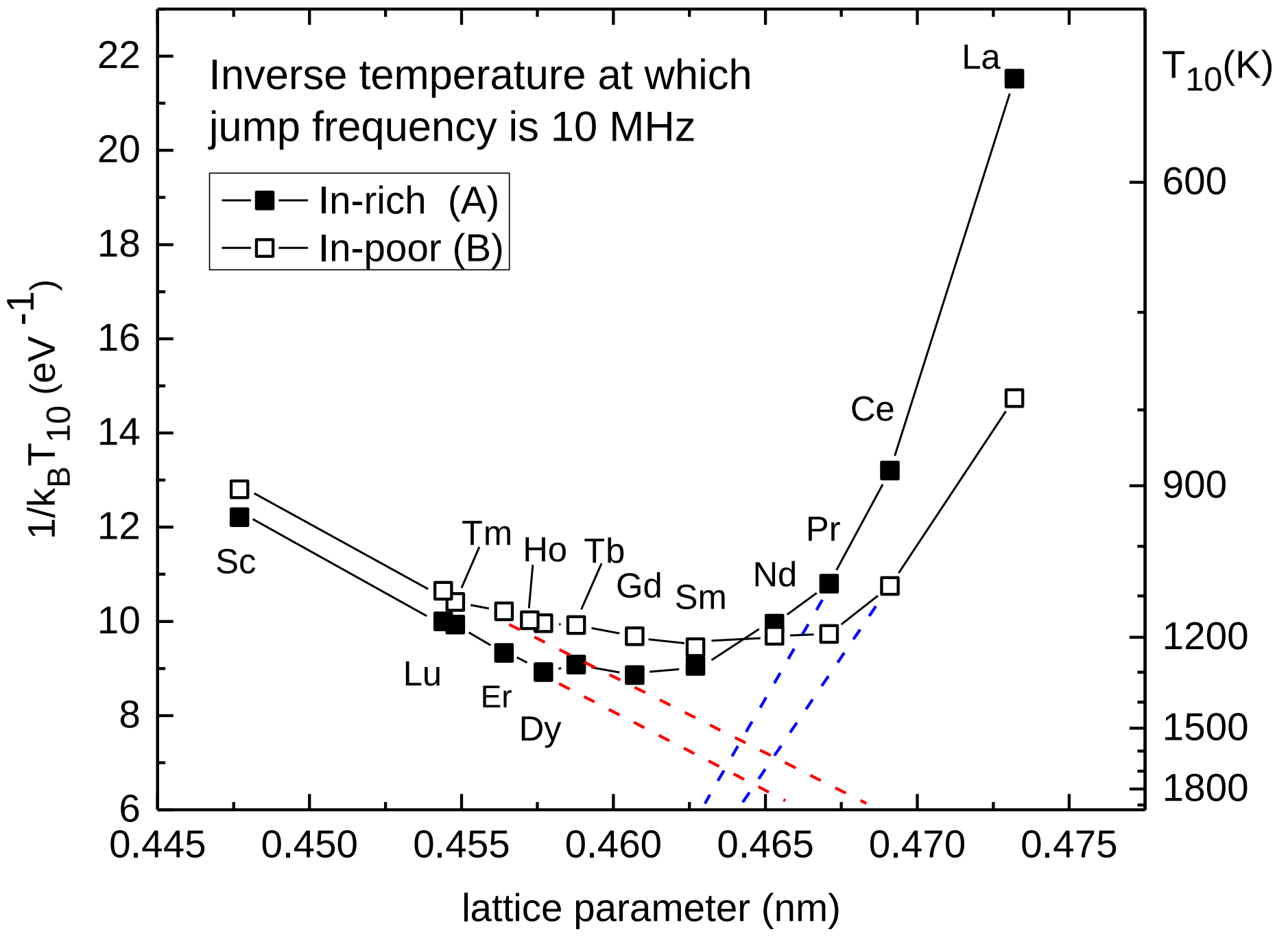

Fig. 4. Inverse temperatures at which jump frequencies of Cd tracer probes in Laves $In_3R$ phases equal 10 MHz. Filled and unfilled symbols show results of measurements for $In_3R$ phases having In-rich and In-poor boundary compositions. The dashed lines show extrapolations of behavior observed at each end of the series, suggesting two different relaxation mechanisms. From [7].

Fitted activation enthalpies were found to be in the range 0.5-2.5 eV and prefactors of order 1-100 THz, which seems consistent with a vacancy diffusion mechanism [9, 10]. Jump-frequency data were fitted using eq. 2 and results for each sample were quantified in terms of the temperature $T_{10}$ at which the jump-frequency in a sample equals 10 MHz. For example, from Fig. 3 it can be seen that for In/Cd probes in $In_3La$, these temperatures are ~580 and 840 K for In-rich and In-poor samples, respectively, from which one can deduce values of $1/k_B T_{10}$ = 21.5 and 14.5 $eV^{-1}$. (Note that a greater value of $1/k_B T_{10}$ implies a greater jump frequency and/or lower activation enthalpy.) The results are plotted in Fig. 4 versus lattice parameters of the phases. Since all the phases are chemically similar, the plot versus lattice parameter serves to reveal a possible influence of lattice parameter on diffusion (e.g. a possible correlation with "lanthanide contraction"). The figure appears to reveal two different diffusion regimes: one for light-lanthanide elements R= La, Ce, Pr and Nd, with large lattice parameters, and the other for heavy-lanthanide elements R= Lu, Tm, Er, Ho, Dy and Tb. Colored dashed lines in Fig 4 are extrapolations of data trends observed from each end of the indide-rare-earth series, and suggest a crossover near the middle of the series between regimes for the different relaxation mechanisms. For the heavy-lanthanide indides, the ratio of jump-frequencies at the two boundary compositions was typically 3, much smaller than for $In_3La$. That is attributed to the ratio of the mole-fractions of $\alpha$-vacancies at the boundary compositions.

The composition dependences of the jump-frequency can be seen in Fig. 4 to be opposite for the two diffusion regimes. Unlike for $In_3La$ and other light-lanthanide elements Ce, Pr and Nd, jump frequencies for In-rich samples for heavy-lanthanide elements (Lu to Tb), and also Y and Sc (filled symbols) are *lower* than for In-poor samples (open symbols). Thus, the behavior for heavy-lanthanide indides is consistent with the $\alpha$-vacancy sublattice diffusion mechanism, since a greater mole fraction of $\alpha$-vacancies exists in In-poor samples than in In-rich ones, leading to a greater jump-rate, as observed.

**Novel explanation for nuclear relaxation in $In_3La$ and other light-lanthanide indides.** A new explanation is developed in this paper for the nuclear relaxation observed in indides of La, Ce, Pr and Nd. The large relaxation frequencies are attributed not to a stationary diffusion mechanism but instead to jumps that transfer daughter Cd-probes from the In- to R-sublattice due to a change in site preference following transmutation of the In-parent probe. $^{111}$In decays by electron-capture with a half-life of 2.8 d to the 417 keV level of $^{111}$Cd, which in turn decays promptly (half-life of 0.12 ns) to the 245 keV PAC level [11]. The observed nuclear relaxation is attributed to non-stationary jumps of the Cd-daughter as an equilibrium distribution of site occupations gets reestablished. Similar behaviors for indides of La, Ce, Pr and Nd suggest a common relaxation mechanism for all. Support for this explanation comes from measurements for $^{111}$In/Cd probes as impurities in the $Pd_3R$ series described below.

**In/Cd probes as impurities in stannides, gallides, aluminides and palladides.** Studies of nuclear relaxation have been carried out for $^{111}$In/Cd probes as impurities in other $L1_2$ series with rare-earth elements (R), including $Sn_3R$ [12], $Ga_3R$ [13], $Al_3R$ [13] and $Pd_3R$ [14, 15]. While $^{111}$In is obviously located on the In-sublattice in $In_3R$ phases at time of decay, lattice locations of impurity probes are not obvious. In addition, site-preferences of dilute impurities in binary phases generally have been found to differ at opposing boundary compositions (see, for example, refs. 6, 16, 17), in the same way as jump frequencies [6, 7]. PAC spectra for impurity probes in $A_3B$ phases (with sublattices labeled $\alpha$ and $\beta$), are superpositions of signals for probes on $\alpha$- and $\beta$-sites, so that the experimental perturbation function is given by

$$G_2(t) \cong f_\alpha \exp(-\lambda t) G_{static}(t) + f_\beta, \tag{3}$$

in which $f_\alpha$ is the fraction of probes on the $\alpha$-sublattice at time of decay of $^{111}$In into $^{111}$Cd, which exhibit a (possibly damped) quadrupolar perturbation function, and $f_\beta$ is the fraction of probes on the $\beta$-sublattice, whose sites have cubic point symmetry and a zero-frequency quadrupole interaction signal that is visible as a vertical offset in the PAC spectrum.

Fig. 5 shows how $1/k_B T_{10}$ varies with lattice parameter for all five series [7], and its examination leads to the following observations:

1. No "universal" dependence of jump frequencies on lattice parameter exists for the five series.
2. As a heuristic rule, dilute impurity solute atoms tend to occupy sites of the sublattice of an element in which there is a deficiency [16]. This leads to a smaller total number of point defects than if the solute were to occupy a site on the other sublattice, thereby forcing creation of an additional antisite defect. All things being equal, the sum of defect formation enthalpies is likely to be smaller when there are fewer point-defects. In accord with the rule, In-impurities were found to primarily occupy $\alpha$-sublattices in R-rich $Pd_3R$, $Al_3R$ and $Ga_3R$ series (i.e., deficient in Pd, Al or Ga), allowing nuclear relaxation caused by jumps on the $\alpha$-sublattice to be observed, with results shown in Fig. 5. Also in accord with the rule, in R-poor samples of those three series, signals were generally observed for probes on $\beta$-sites (an unperturbed quadrupolar interaction attributed to $^{111}$In/Cd probes on cubic sites of the rare-earth $\beta$-sublattice, not shown), so that diffusional relaxation for probes on the $\alpha$-sublattice was unobservable owing to a small site fraction. As an exception, In-probes partially occupied $\alpha$-sublattices for both R-rich and R-poor compositions of $Sn_3La$ and $Sn_3Ce$. Thus, there was a relatively stronger site-preference for In to occupy Sn-sites than R-sites in those phases. It can be seen from Fig. 5 that the jump-frequency was greater for the Sn-poor samples in those two Sn-phases, opposite to what was observed for R= La, Ce, Pr and Nd indides and consistent with diffusion taking place via the simple $\alpha$-sublattice diffusion mechanism. Another exception was $Pd_3La$, for which no probes were observed to occupy Pd-sites in either Pd-rich or Pd-poor samples, indicating a strong preference for In-impurities to occupy the $\beta$-site, regardless of composition.

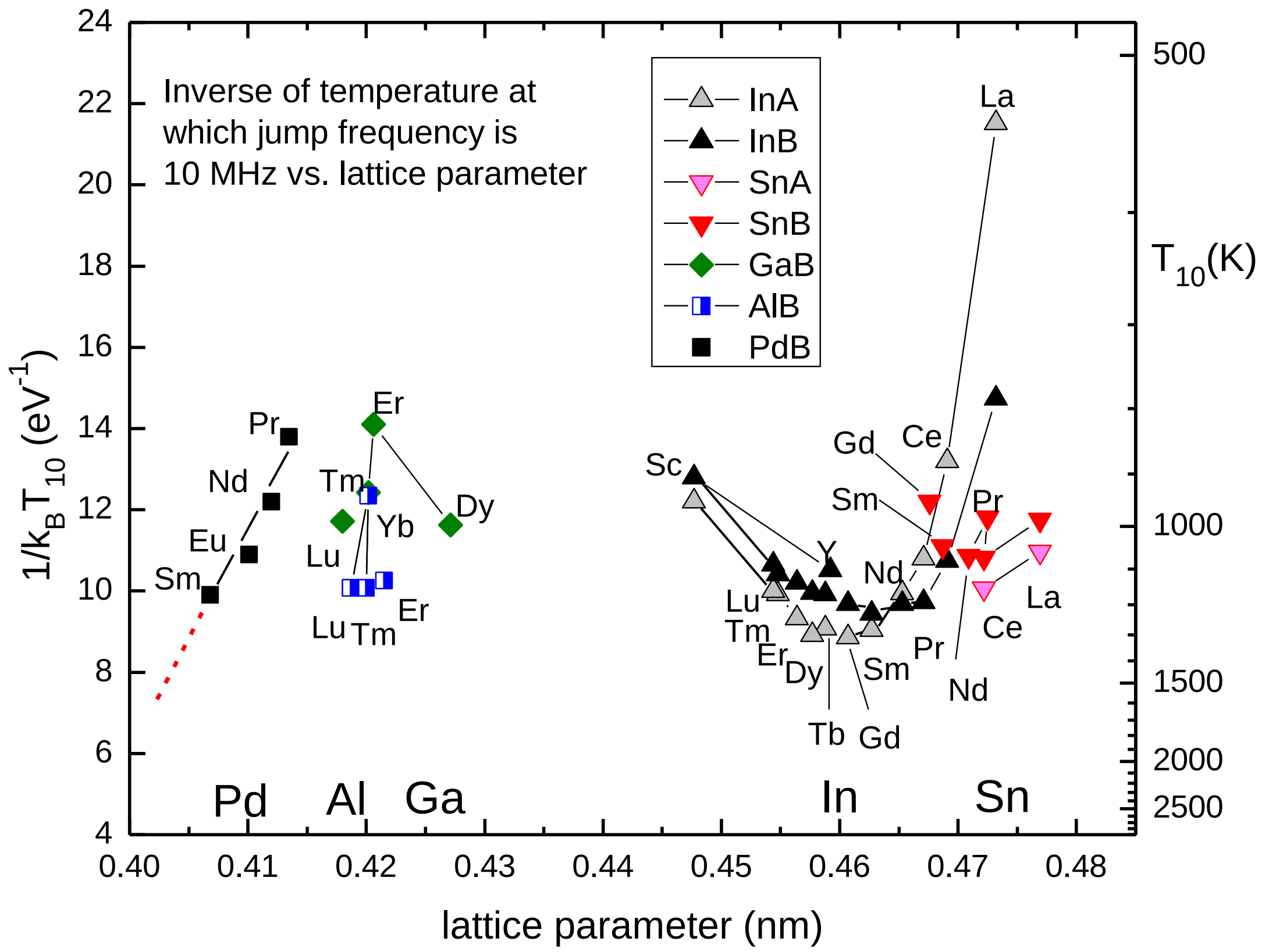

Fig. 5. Inverse temperatures at which jump frequencies of Cd tracer probes in Laves $In_3R$, $Sn_3R$, $Ga_3R$, $Al_3R$ and $Pd_3R$ series equal 10 MHz (from [14]). Filled and unfilled symbols show results of measurements on phases having R-poor (A) and R-rich (B) boundary compositions. For series such as $Pd_3R$, Cd-probes only occupied Pd-sites in samples having Pd-poor boundary compositions, so that measurements of nuclear relaxation could not be made for Pd-rich samples.

3. Trends in the light-lanthanide indides and palladides are similar, with parameters $1/k_B T_{10}$ for both series climbing rapidly along the sequence of light lanthanide elements from Nd to Pr, Ce, and La. This is considered to suggest a common origin. As noted above, stannides have the opposite composition dependence of the relaxation frequency and do not exhibit the same rapid climb in the $1/k_B T_{10}$ parameter for the four light-lanthanide elements as observed in the indides and palladides. Data for the aluminides and gallides are too fragmentary to exhibit any clear trend.

**Experiments on $^{111}$In/Cd probes in $Pd_3R$ phases.**

Qiaoming Wang carried out an extensive study of site-preferences of $^{111}$In probes and jump-frequencies of $^{111}$Cd daughter probes in rare-earth palladides $Pd_3R$, all of which have the $L1_2$ structure [15]. His thesis and an addendum containing additional information and analysis compiled subsequently by Randal Newhouse and myself are available in a research report [18]. Results of this research have appeared partially in [14].

PAC spectra measured on Pd-poor samples of two phases will be used to illustrate the results [14]. In Figs. 6 and 7 are shown PAC spectra for $Pd_3Eu$ and $Pd_3Pr$ measured at the indicated temperatures (from Figs. 2 and 4 in [14]). The "double-sided" spectra shown in Figs 6 and 7 exhibit equivalent, independent data at nominal negative time that mirrors the data at positive time following creation of the PAC level at time zero, in the middle of the spectrum. They are particularly useful for examining the shape of the perturbation function close to time-zero. For $Pd_3Eu$ (Fig. 6), the time-domain spectrum at 673 K is dominated by a periodic 55-ns quadrupolar perturbation attributed to probes on Pd-sites and a small vertical offset attributed to probes on cubic Eu-sites. With increasing temperature, the vertical offset increases at the expense of the periodic signal. This is attributed to a thermally activated transfer of probes from Pd-sites to Eu-sites. With increasing temperature, the periodic signal also becomes more damped, with the perturbation appearing as the product of a static perturbation (such as seen at 673 K) and an exponentially decaying envelope function attributed to fluctuating electric-field gradients (see eq. 3 and compare with spectra in Fig. 2). If the relaxation were caused by the simple $\alpha$-sublattice diffusion mechanism, the decay time of the perturbation would be equal to the inverse of the mean jump-frequency of the probe. From the observed decay time of ~200 ns in the spectrum at 1023 K, one would infer a diffusional jump-frequency of 5 MHz. A different cause is proposed below for the observed relaxation.

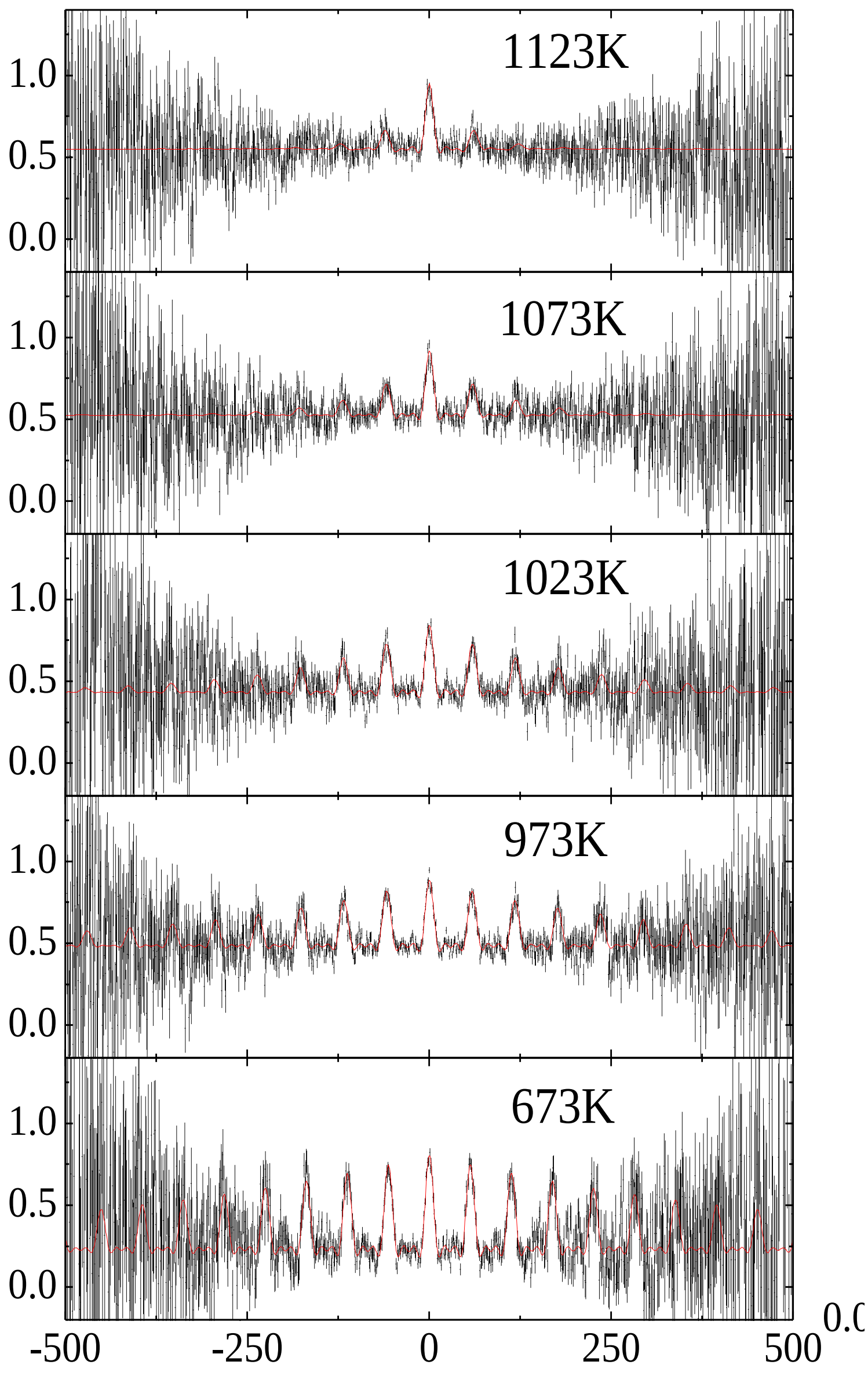

Fig. 6. PAC spectra of $Pd_3Eu(^{111}In/Cd)$ at the indicated temperatures. Quadrupole perturbations with zero frequency and with a period of 55 ns are attributed to probe atoms at time zero on the Eu- and Pd-sublattices, respectively. Damping of the 55-ns signal at high temperature is attributed to nuclear relaxation due to stocastic jumping of the daughter Cd-probes. Note that the site fraction (amplitude) of the zero-frequency signal increases at the expense of the precessional signal with increasing temperature. From [14].

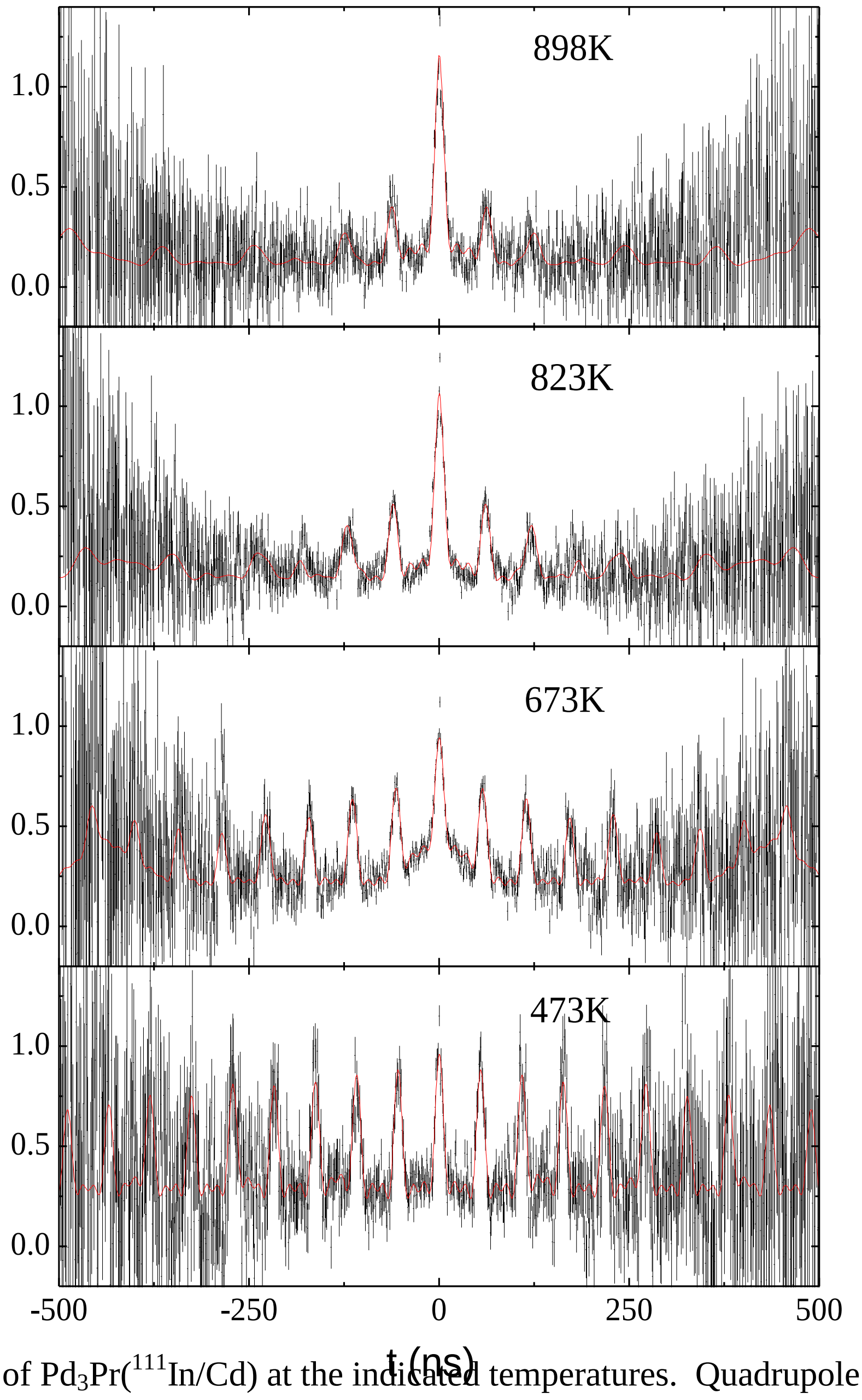

Fig. 7. PAC spectra of $Pd_3Pr(^{111}In/Cd)$ at the indicated temperatures. Quadrupole perturbations with zero frequency and with a period of 55 ns are attributed to probe atoms at time zero on the Pr- and Pd-sublattices, respectively. Damping of the 90-ns signal at high temperature is attributed to nuclear relaxation due to stochastic jumping of the daughter Cd-probes. The site fraction of the zero-frequency signal decreases at the expense of the precessional signal with increasing temperature. From [14].

In Fig 7 are PAC spectra measured for Pd-poor $Pd_3Pr$. The spectra again show a periodic perturbation but also an offset that is larger at low temperature, the opposite of what was observed for $Pd_3Eu$. With increasing temperature, the amplitude of the periodic signal grows, as evidenced in particular by the height of the peak close to time-zero, while the amplitude of the vertical offset, attributed to probes on cubic Pr-sites, decreases. This indicates a thermally-activated shift in site-preference of In-probes from Pr-sites at low temperature to Pd-sites at high temperature, a behavior opposite to that observed for $Pd_3Eu$ in Fig. 6. At the same time, there is a large increase in nuclear relaxation with increasing temperature.

Spectra for the other palladide phases were analyzed in the same way [14, 15, 18]. An overview of results for all $Pd_3R$ phases is provided in tabular form in Fig. 8. Blank entries indicate either that no measurements were made at all (R= Pm, Gd, Dy, Ho, Tm, and R-poor Y and Sc) or that jump-frequencies for probes on the $\alpha$-sublattice in Pd-poor samples were too low to be detected in measurements up to 1100 K (R= Tb, Er, Yb, Lu, Y, Sc).

**Site-preferences (Fig. 8a).** For the R-poorer boundary compositions (bottom row), there was a strong site-preference of $^{111}$In probes for cubic R-sites and no signals were detected in any of the R-poor phases for probes on the $\alpha$-sublattice. For Pd-poorer compositions (top row), there was a monotonic trend along the series (also monotonic in lattice parameters), as follows:

R= Sc, Y, Lu, Yb, Er and Tb. There was a strong site-preference for probes to occupy the Pd-sublattice, with no probes detected on R-sublattices. No diffusional relaxation was detected in measurements up to 1100 K. (See, e.g., spectra for $Pd_3Lu$ in Fig. 1 of [14]).

R= Sm and Eu. Pd-sites were preferred by the In-probes at low temperature, with transfer to R-sites with increasing temperature. (Confer Fig. 6.) The spectra exhibit diffusional relaxation of the quadrupole interaction for probes on the Pd-sublattice.

R= Nd, Pr and Ce. Compared with R= Sm and Eu, the opposite change in site-preference was observed, with the site-fraction of probes on R-sites decreasing with increasing temperature (confer Fig. 7.) The spectra exhibited significantly greater nuclear relaxation.

R= La. The site-preference towards R-sites was strong at both boundary compositions and no probes were detected on Pd sites.

**Jump-frequencies (Fig. 8b).** Measurements of relaxation could only be made on Pd-poor samples. Although site fractions of probes on Pd-sites in Pd-poor samples were close to 100% for Pd-poor R= Sc, Y, Lu, Yb, Er, and Tb samples, no relaxation was observed up to 1100 K. Increasing relaxation was observed along the sequence R= Sm, Eu, Nd, Pr and Ce. Jump frequencies could not be measured for R= La owing to a strong site-preference for In-probes to occupy La-sites in both Pd-rich and Pd-poor samples.

**Correlation between In site-preferences and Cd jump-frequencies in the $Pd_3R$ series.**

In thermal equilibrium the ratio of site fractions of In-parent probes on the two sublattices of an $A_3B$ phase is thermally activated according to

$$\frac{f_\beta}{f_\alpha} = A\exp(-Q_{transf}/k_B T), \qquad (4)$$

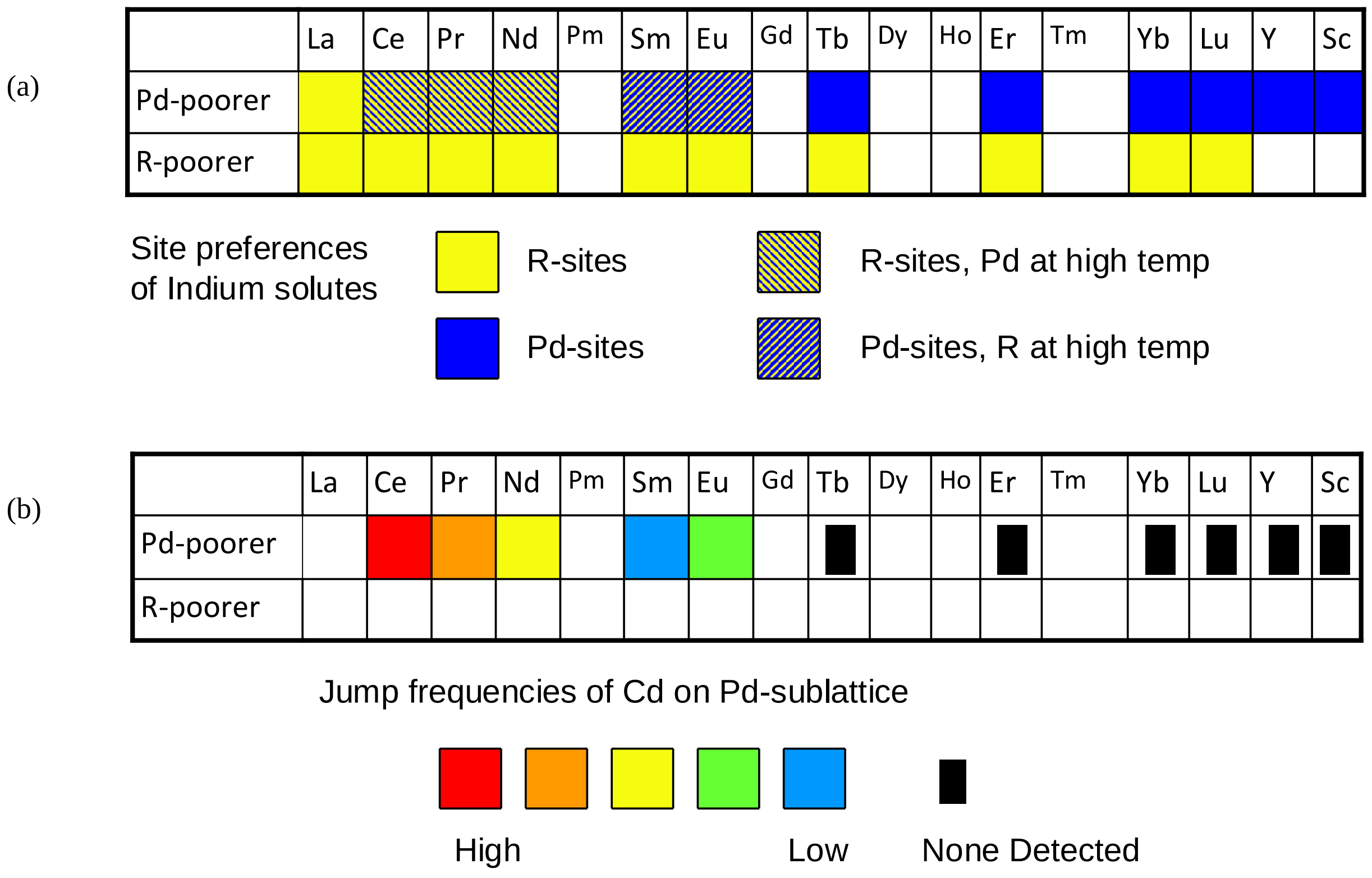

Fig. 8. Overview of results of measurements in $Pd_3R$ phases for (a) site-preferences of $^{111}$In parent probes and (b) jump-frequencies of $^{111}$Cd daughter probes. The two tables summarize results for measurements on pairs of samples having Pd-poorer and R-poorer boundary compositions. Blank entries in (a) indicate no measurements. Blank entries in (b) indicate either no measurements or that jump-frequencies were too low to be detected for probes on the $\alpha$-sublattice in measurements (jump frequency less that 1 MHz at 1100 K.)

in which $f_\beta$ and $f_\alpha$ are site fractions of In probes on the R- and Pd-sublattices, $A$ is a prefactor and $Q_{transf}$ is an activation enthalpy for transfer of probes between the two sublattices. Such activation enthalpies have been measured previously in this laboratory for $^{111}$In probes in $GdAl_2$ [17], $Ga_7Pd_3$ [19], and $Al_3Ti$ and $Al_3Zr$ [20]. Experimentally determined values of $Q_{\text{transf}}$ are shown for R= Ce, Pr, Nd, Sm, and Eu phases in Fig. 9. These were obtained from fits of Arrhenius plots of experimental site-fraction ratios with eq. 4. It can be seen that $Q_{\text{transf}}$ changes sign in the series between Nd and Sm, consistent with the qualitative observations made above comparing evolutions with temperature of spectra for $Pd_3Eu$ and $Pd_3Pr$. For other heavy-lanthanide phases studied, R= Tb, Er, Yb, Lu, and Y and Sc, site-fractions of probes on the R-sublattice were too small to estimate the site-fraction ratio with any precision. For R= La, the fraction of probes on the Pd-sublattice was too small to estimate the ratio. In summary, it can be seen that the In-impurity has a monotonically changing site-preference from strongly favoring Pd-sites at the Lu-end of the lanthanide series to favoring R-sites at the La-end. The data show that the ground-state site of In in Ce, Pr and Nd palladides is the R-site, with the site fraction of probes on Pd-sites increasing with temperature. The opposite holds true for Sm and Eu palladides in which the ground state is the Pd-site, with the site-fraction on R-sites increasing with temperature. These trends are what one expects since the site-preference should become more random at high temperature. Importantly, it will be argued below that the daughter Cd-probe has a similar monotonic change in site-preference along the series owing to the chemical similarity of sp-elements In and Cd.

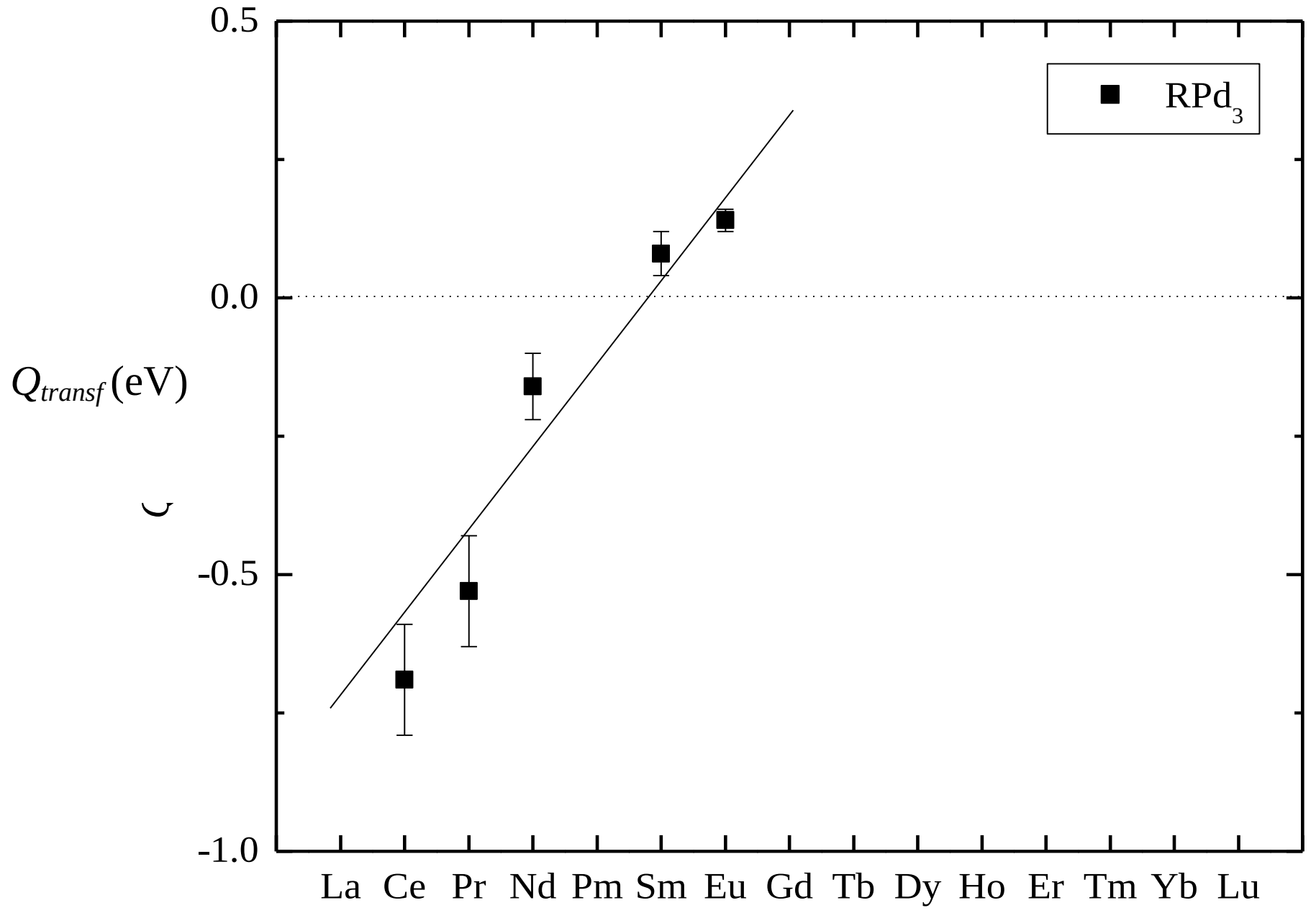

Fig. 9. Dependence of the site-fraction ratio activation enthalpy $Q_{transf}$ on atomic number in $RPd_3$ phases. From an addendum to the MS thesis of Q.Wang [18].

The nuclear relaxation was also obtained from fits of PAC spectra for Pd-poor $Pd_3R$ phases using eq. 3. Taking the jump-frequency *w* to be equal to the fitted relaxation frequency $\lambda$, its temperature dependence was fitted with the thermally activated expression given in eq. 2. The five members of the series (R= Eu, Sm, Nd, Pr, Ce) in which nuclear relaxation of the Cd-probe was observed are coincidently the five in which the site-preference of the parent In-probe was observed to change. Arrhenius plots of the temperature dependences of *w* are shown for the five phases in Fig. 10. It is can be seen that the activation enthalpies, given by slopes of the data sets, decrease precipitously along the series R= Sm, Eu, Nd, Pr, Ce. This trend is made clearer by plotting the fitted activation enthalpies versus lattice parameters in Fig. 11. The figure shows a linear trend in activation enthalpies, with *Q* decreasing from 2.2 eV for $Pd_3Sm$ to 0.2 eV for $Pd_3Ce$.

From Fig. 10 it can be seen that the relaxation frequency at 1000 K is a factor of ~$10^3$ larger in $Pd_3Pr$ than in $Pd_3Sm$. It is difficult to interpret such an enormous difference in terms of the $\alpha$-sublattice vacancy diffusion mechanism since it would suggest also a thousand-fold difference in vacancy mole fractions in two phases of very similar chemistry.

To recapitulate the main experimental results:

1. Indium has a dramatic change in site-preference along the palladide series from R=Lu to La.
2. From Lu to Tb, there is a strong site-preference of In for the Pd-site, and no observed nuclear relaxation of the quadrupole perturbation for $^{111}$Cd-probes that originate on the Pd-sublattice.
3. From Eu to La, the preferred site of In in Pd-poor samples changes progressively from Pd- to R-sites. At the same time, the relaxation frequency of daughter Cd-probes starting on Pd-sites increases by three orders of magnitude.
4. A strong negative correlation exists between activation enthalpies $Q_{transf}$ and *Q* shown in Figs. 9 and 11, but it must be remembered that they refer to different probes, the parent In and daughter Cd, respectively. However, it makes sense that the relaxation frequency of the Cd-probe should increase as its stability on the Pd-sublattice (presumed to be similar to that of In) decreases.

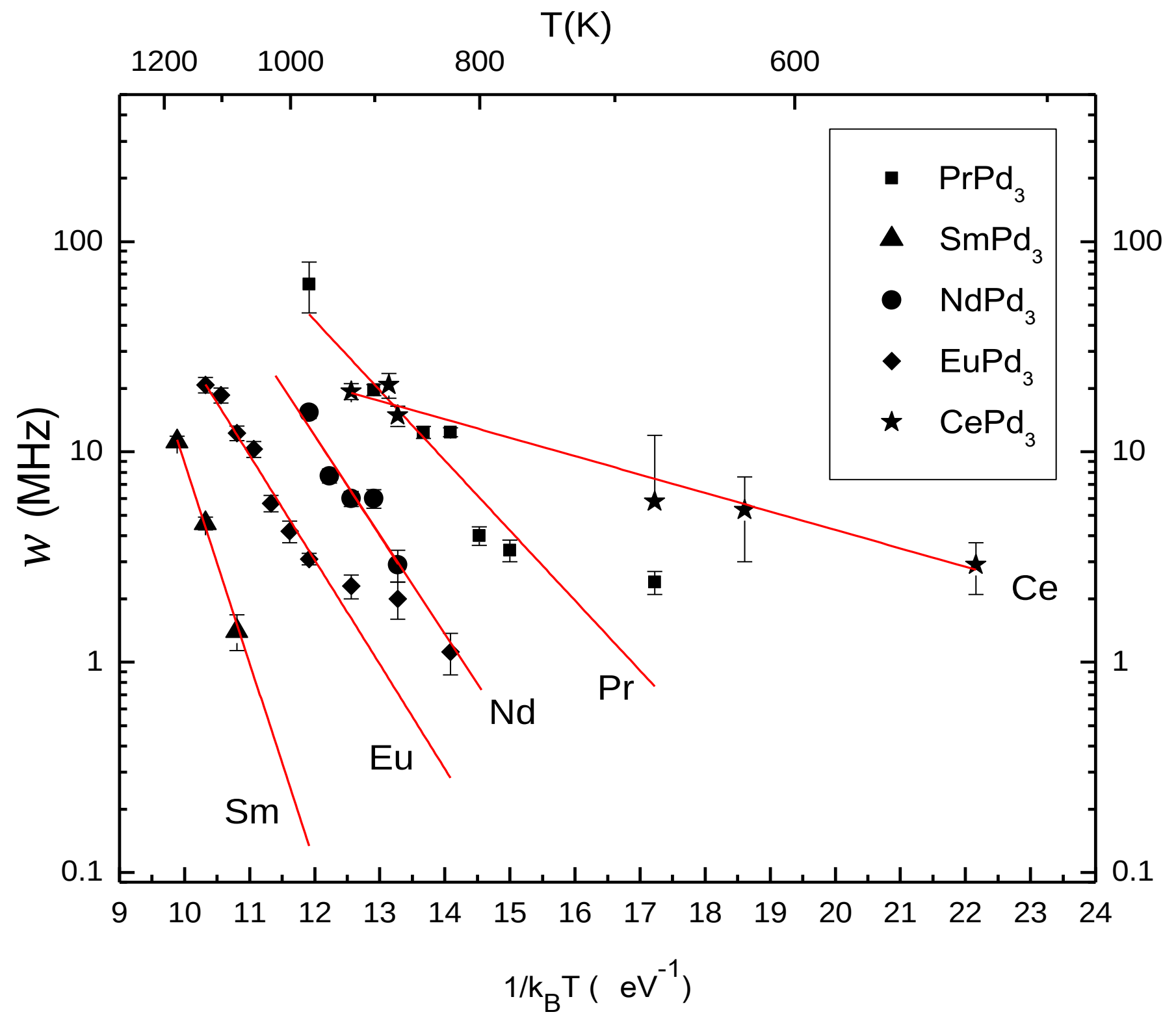

Fig. 10. Arrhenius plots of jump-frequencies of In/Cd probe atoms in five $RPd_3$ phases. From an addendum th the MS thesis of Q. Wang [18].

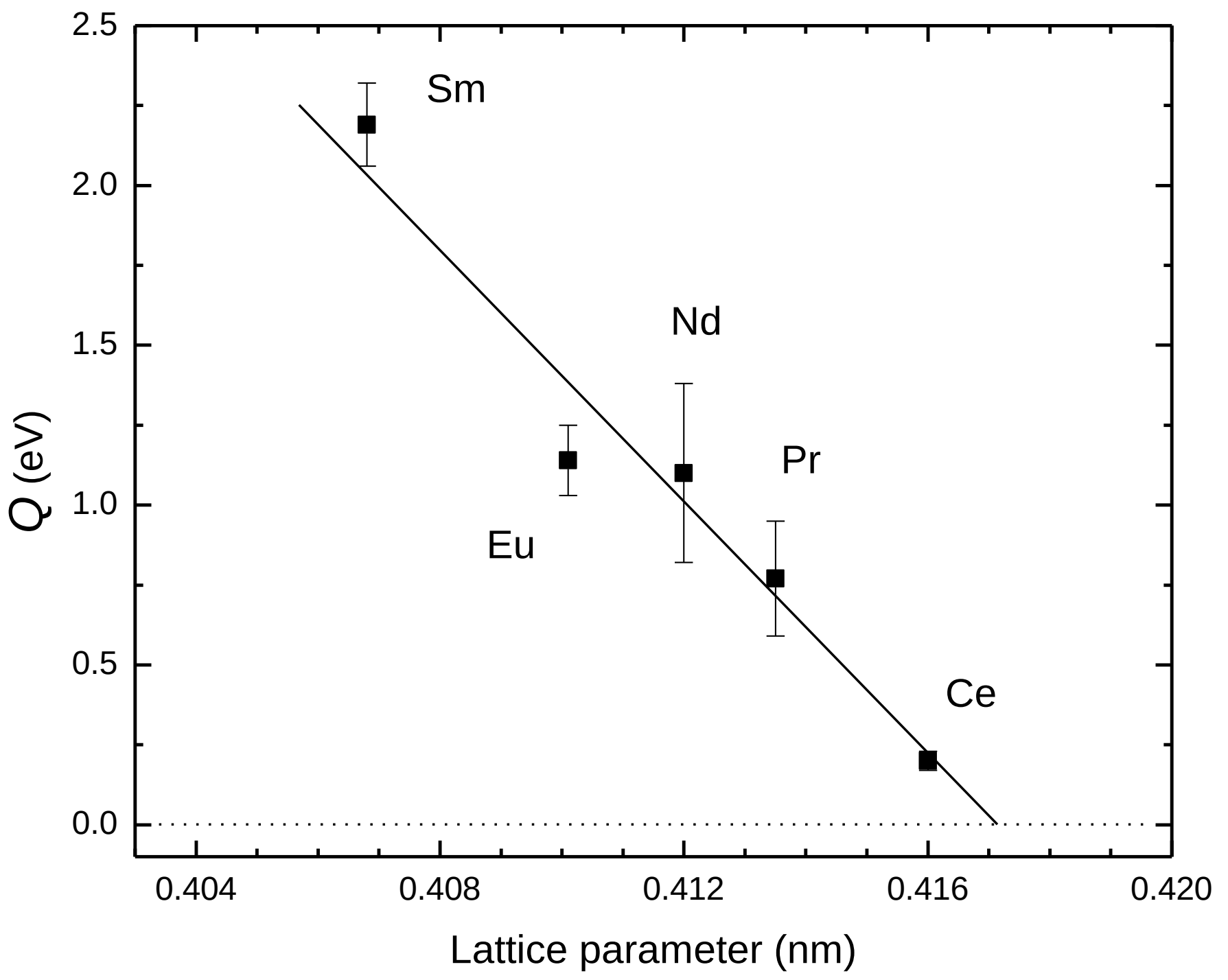

Fig. 11. Dependence on lattice parameter of the jump-frequency activation enthalpy for In/Cd probes in $RPd_3$ phases. From an addendum to the MS thesis of Q. Wang [18].

**Origin of nuclear relaxation in the light-lanthanide indides and palladides.** Nuclear relaxation in both series of phases is now reinterpreted in terms of an approach to a new equilibrium distribution of daughter probes following radioactive decay.

1. Site-preferences of Cd- and In-probes in $Pd_3R$ phases are assumed to change in a similar way along the lanthanide series of Pd-phases due to the likeness in chemistry and atomic volume of these neighboring *sp*-elements. Just as for In-probes (Fig. 8a), Cd impurities are assumed to have a strong site-preference for Pd-sites in the heavy-lanthanide palladides (R= Lu, Tm, Er and Dy), weakening and turning into a strong preference for R-sites along the light-lanthanide series R= Eu, Sm, Nd, Pr, Ce and La. The increasing relaxation frequency of the Cd-probe in light-lanthanide palladides with R= Eu, Sm, Nd, Pr, and Ce is taken as a measure of the increasing instability of the probe's position on Pd-sites.

2. For heavy-lanthanide indides, the observed composition dependence of the nuclear relaxation is consistent with the $\alpha$-sublattice vacancy mechanism (see Fig. 4 and discussion). Jump frequencies were relatively low and changes in jump-frequencies or activation enthalpies are small from one element to the next. Information about diffusion mechanisms from composition dependences is unavailable for the palladides because site-preferences of In-impurities precluded measurements for probes on $\alpha$-sites in any Pd-rich samples.

3. For light-lanthanide indides, the observed composition dependence is inconsistent with the simple $\alpha$-sublattice diffusion mechanism and, moreover, the relaxation grows very rapidly in the sequence R= Nd, Pr, Ce to La, as shown in Figs. 10 and 11, unlike the relatively smaller changes observed for heavy-lanthanide indide series in Fig. 4. The relaxation likewise grows rapidly in the light-lanthanide palladides, and they are presumed to have a common origin. This implies that daughter Cd-probes remain on In- or Pd-sites in heavy-lanthanide phases but transfer to R-sites in light-lanthanide phases following transmutation of In-probes.

4. Activation enthalpies for nuclear relaxation in the light-lanthanide phases are extraordinarily low, $Q$= 0.53 or 0.81 eV for $In_3La$ (Fig. 3) and 0.8 eV for $Pd_3Ce$ (Fig. 11). The melting temperature of $In_3La$ is $T_m$= 1400 K, from which one obtains a ratio $Q/k_BT \cong 4.4$, far below comparable values of activation enthalpies for diffusion observed in metals and intermetallic phases [21]. From Fig. 4 it can be seen that a relaxation rate of 10 MHz attributed to the simple $\alpha$-sublattice diffusion mechanism typically occurs in the heavy-lanthanide indides at a temperature of ~1250 K. For $In_3La$, the same relaxation rate occurs at a much lower temperature of 500-700 K and, as noted previously, extrapolation of the upper curve in Fig. 3 gives a relaxation frequency of ~1 kHz at room temperature.

**Possible transfer mechanisms.** The high relaxation frequencies and low activation enthalpies observed for light-lanthanide palladide and indide phases suggest that equilibration of site-preferences occurs very rapidly following radioactive decay. Observation of a greater relaxation frequency for In-rich samples than for In-poor ones suggests that the relaxation mechanism is mediated by R-vacancies and not $\alpha$-vacancies. Five alternative transfer mechanisms are considered for In- or Pd-rich samples:

(1) The most elementary transfer of $Cd_\alpha$ to $\beta$-sites would be by $Cd_\alpha$ jumping into an adjacent $V_\beta$;

$$Cd_\alpha + V_\beta \rightarrow Cd_\beta + V_\alpha . \tag{5}$$

Presumably, such an exchange can occur between $Cd_\alpha$ on an $\alpha$-site and $V_\beta$ on any of the probe's near-neighbor $\beta$-sites, but not with more distant vacancies. Since no inhomogeneous broadening has generally been observed in spectra of $In_3R$ phases, mole fractions of vacancies or other point defects are believed to be low, say of order $10^{-3}$ or smaller. This makes it implausible that the reaction in eq. 5 could equilibrate site-preferences of all Cd-probes since only probes next to a $V_\beta$ could transfer. Consequently, one of the reacting species in eq. 5 must undergo long-range migration in order to form reaction pairs for all Cd-probes.

(2) An R-vacancy might undergo long-range migration and come next to a $Cd_\alpha$ defect, enabling the two to react via eq. 5. Since R-sites are isolated and do not form a connected sublattice, $V_\beta$ cannot migrate by near-neighbor jumps that do not create additional point defects. $V_\beta$ might diffuse by sequences of six near-neighbor jumps [22], but the activation enthalpy barrier for six-jump cycles is expected to be high, owing to the need to create three additional, transient point defects in the course of a cycle. This makes it improbable that activation enthalpies as low as 0.53 eV could be explained by such a mechanism.

(3) The $Cd_\alpha$-probe could diffuse on the majority sublattice via the $\alpha$-sublattice vacancy diffusion mechanism until it comes next to a $\beta$-vacancy and then react via eq. 5. But diffusion would then be mediated by $\alpha$-vacancies that are present in only low mole fractions in the In- or Pd-rich samples, leading to low migration rates. The strong composition dependence observed in the light-lanthanide indides appears to rule this mechanism out (compare data sets shown in fig. 4).

(4) $Cd_\alpha$ might replace an adjacent R-atom by kicking it into a neighboring $V_\alpha$:

$$Cd_\alpha + R_\beta + V_\alpha \rightarrow V_\alpha + Cd_\beta + R_\alpha \, . \tag{6}$$

In such a "kick-out" reaction, Cd transfers to the $\beta$-sublattice with creation of an antisite $R_\alpha$ defect while regenerating another $V_\alpha$. This reaction would be favored if the sum of formation enthalpies of the three defects on the right-hand side of eq. 6 is less than of the two on the left-hand side. However, the mechanism obviously requires presence of $\alpha$-vacancies, so that the reaction rate should again be greater in In-poor alloys and not, as observed experimentally, in In-rich alloys.

(5) Finally, in a variation of mechanism (1) above, $Cd_\alpha$ might dissociate to an interstitial position from its original $\alpha$-site, diffuse interstitially, and then trap at $V_\beta$ to form a transferred $Cd_\beta$ defect:

$$Cd_\alpha \rightarrow V_\alpha + Cd_i \, , \tag{7a}$$
$$Cd_i \rightarrow Cd_i \, , \tag{7b}$$
$$Cd_i + V_\beta \rightarrow Cd_\beta \, . \tag{7c}$$

Since activation enthalpies for interstitial diffusion are generally low, the Cd-interstitial might migrate at temperatures lower than those typical for vacancy diffusion. Owing to the very low mole fraction of Cd-defects (< $10^{-11}$), mole fractions of $V_\alpha$ and $V_\beta$ would be insignificantly affected by this solute transfer. The transfer can also proceed in In-poor samples as long as the minority mole fraction of $V_\beta$ remains greater than the extremely low mole fraction of $Cd_\alpha$. This mechanism is consistent with the observed composition dependence of jump-frequencies in the light-lanthanide indides and seems to allow for transfer at low temperatures. The reaction steps in eq. 7 are pictured together schematically in Fig. 12.

**Interstitial transfer mechanism.** Mechanism (5) appears most consistent with the observations. It is reasonable to suppose that the interstitial $Cd_i$ traps irreversibly at $V_\beta$ (reaction 7c). Presumably, the activation enthalpy for formation of the interstitial (7a) is lower for $In_3La$ than for R= Ce, Pr or Nd owing to its larger lattice parameter and a more open structure that can accommodate the interstitial Cd-defect. The 0.53 eV activation enthalpy observed for In-poor $In_3La$ (Fig. 3) might be roughly equal to the activation enthalpy for reaction (7a). The decrease in the relaxation rate in the indides along the series R= La, Ce, Pr and Nd could be a consequence of an increase in the formation enthalpy of the interstitial owing to shrinking of the lattice parameter that increases steric hindrance. The greater relaxation rate in In-rich than in In-poor $In_3La$ may be due to an increased chemical potential for In that serves in effect to "eject" solutes from the In-sublattice.

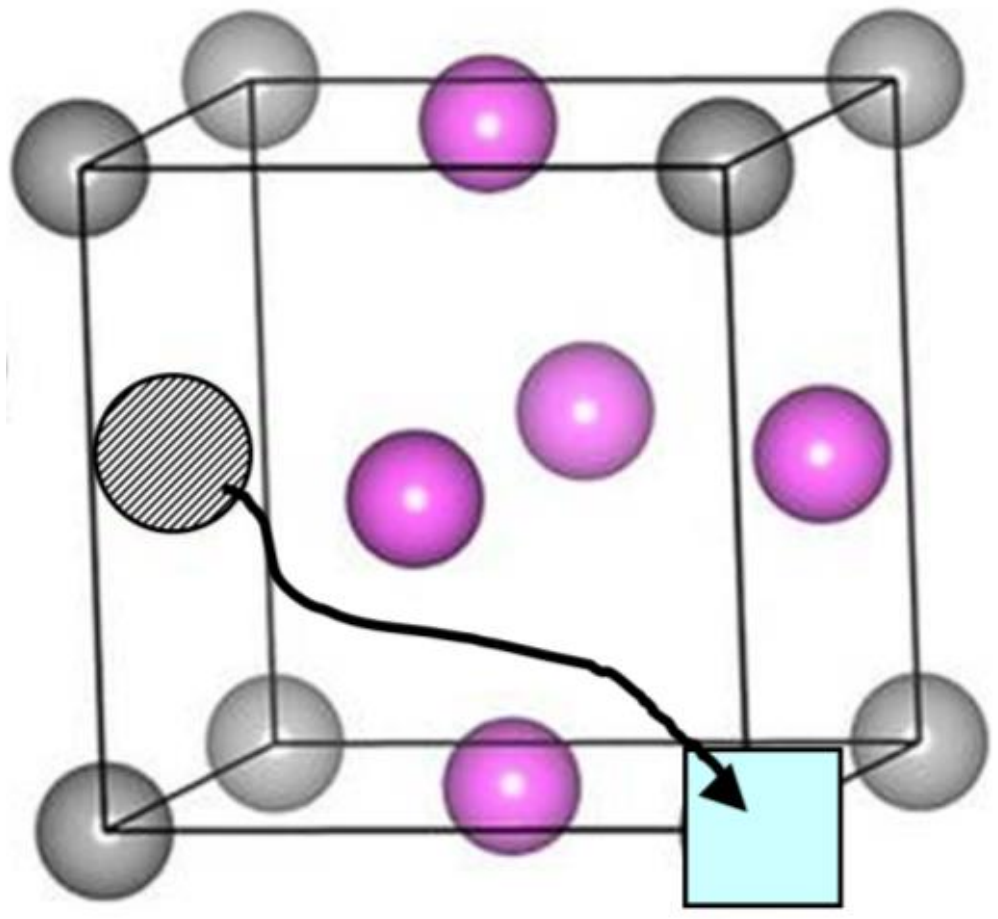

Fig. 12. Interstitial transfer mechanism (eq. 7). Following decay of $^{111}$In into Cd, the Cd probe is ejected from a site on the $\alpha$-sublattice, diffuses interstitially, and then traps at a vacant $\beta$-site.

**Discussion.**

**Decomposition of the two relaxation regimes in the lanthanide indides.** The curves in Fig. 4 delineate two diffusion regimes, one dominant for the heavy-lanthanides and the other for the light-lanthanides. Extrapolations of either regime into the domain of the other are indicated by the drawn dashed lines. It can be seen that the interstitial transfer mechanism dominates for R= Nd, Pr, Ce and La and the $\alpha$-sublattice vacancy diffusion mechanism for R= Lu, Tm, Er, Ho, Tb and Gd (and also Sc). Taking the dashed-line extrapolations at face value leads to the following observations. (1) The jump-frequency for $\alpha$-sublattice vacancy diffusion of Cd probes decreases rapidly as the lattice parameter increases along the series R= Sc, Lu, Tm, Er, Ho, Dy, Tb and Gd. For example, the temperature at which the jump frequency equals 10 kHz is 20% lower for Lu than for Dy. This is a significant and very interesting trend, for which a clear explanation is lacking. It might be caused by an increase in the effective activation enthalpy for formation of $\alpha$-vacancies or in the migration enthalpy of $\alpha$-vacancies along the series. (2) Nuclear relaxation attributed to interstitial transfer of Cd-solutes increases dramatically along the series R= Nd, Pr, Ce and La, and is completely dominant for R= Pr, Ce, and La.

**Relaxation regimes in the lanthanide palladides.** In contrast to the indides, no diffusional relaxation was observed up to 1100 K for probes in heavy-lanthanide Pd-poor palladides. It is assumed that diffusion via the $\alpha$-sublattice vacancy diffusion mechanism occurs, but at rates too low to be detected (the threshold frequency for detection of relaxation is ~1 MHz using PAC of $^{111}$In/Cd, which is a function of the lifetime of the PAC level). Relaxation was only observed among light-lanthanide palladides in Pd-poor samples with R= Sm, Eu, Nd, Pr and Ce. The relaxation rates are comparable to rates for members of the same series in the In-poor indides and are also attributed to interstitial transfer of daughter Cd-probes from $\alpha$- to $\beta$-sublattices.

**The nature of nuclear relaxation in the light-lanthanide phases.** In the slow-fluctuation regime, the dynamically relaxed perturbation function appears as the product of a static quadrupole perturbation function and envelope function $\exp(-\lambda t)$ which, when fitted, determines the relaxation frequency $\lambda$ (eq. 1). For stationary diffusion, for example via the $\alpha$-sublattice vacancy diffusion mechanism, the relaxation frequency is equal to the mean jump frequency $w$ ($\lambda \cong w$), which represents the inverse of the mean residence time of the Cd-probe on an $\alpha$-site before it jumps to another $\alpha$-site. For the interstitial transfer mechanism, a nonstationary process reestablishing thermodynamic equilibrium of site-occupations after nuclear decay, there is no homogeneous, repeated sequence of jumps. Instead, there is a first jump of the probe to an interstitial site, with change in EFG, followed by an indeterminate number of additional jumps while the probe diffuses interstitially, with additional but different changes in EFGs, and terminating in a final jump onto a $\beta$-site. While it is impossible to model $\lambda$ quantitatively, it is reasonable to suppose that the time of the first jump and magnitude of the first change in EFG are by far the most important. As a crude approximation, one can assume that the change in EFG will be comparable to that of a jump in the $\alpha$-sublattice diffusion mechanism. It then follows that the mean time to the first jump after transmutation will be given approximately by the inverse of the fitted relaxation frequency, perhaps asserting a systematic uncertainty by a factor of 4 to account for lack of knowledge of the magnitude of the EFG changes in the first and subsequent jumps. Thus, Fig. 3 for $In_3La$ needs to be newly interpreted as an Arrhenius plot of the inverse times until the first jump of the daughter Cd-probe from its original location on the In-sublattice. For example, the time to the first jump in In-rich $In_3La$ is 1 ns at 1000 K and, from an extrapolation, 1 ms at room temperature. Other values can be inferred from data for R= Pr and Ce in Figs. 4 and 10 and tabulated in refs. [9, 10].

**Density-function theory (DFT) calculations to test observations and hypotheses.** DFT calculations using programs such as WIEN2k [[23]] can test various experimental observations and hypotheses presented in this paper. These include the observed change in site-preferences of In solutes along the series of $Pd_3R$ phases and the inferred changes in site preference for Cd-solutes in the palladides and indides. The enthalpy of formation of a Cd-interstitial via reaction 7a could be calculated and compared with measured activation enthalpies such as 0.53 eV for the relaxation process in In-rich $In_3La$.

**Relaxation experiments using other PAC probes.** An important feature of experiments using the $^{111}$In/Cd probe is prompt decay of the parent probe to the PAC level of the daughter. The situation is quite different for the second most commonly used PAC probe, $^{181}$Hf/Ta. $^{181}$Hf undergoes beta decay with a half life of 42.4 d, mostly populating the 615 keV excited state of $^{181}$Ta that has a lifetime of 17.8 $\mu$s [11]. The 615 keV level subsequently decays to the 482 keV PAC level, and then to the ground state. The long lifetime of the precursor 615 keV level gives ample time for daughter Ta-probes to reach an equilibrium population distribution, due to a change in site preference between parent and daughter probe, prior to formation of the PAC level. For the $^{181}$Hf/Ta decay scheme, an interstitial transfer mechanism like that proposed here would go unobserved. On the one hand, equilibration ensures that only true diffusional relaxation is observable. On the other hand, transient transfer phenomena such as those considered here cannot be studied. A related aspect is the initial lattice location of the PAC probe atom. With prompt decay to the PAC level of $^{111}$Cd, one can be certain that the Cd-probes in $In_3R$ start life on In-sites. With the long delay preceding formation of the PAC level in $^{181}$Ta, one cannot predict the starting location of the daughter probe with any certainty. A similar situation to that for the $^{111}$Hf/Ta probe applies, of course, for isomeric PAC probes such as $^{111m}$Cd/Cd.

**Nuclear relaxation of probes starting on cubic lattice sites.** It is interesting to consider if motional relaxation can be used to measure jump-frequencies of impurity probes that start on cubic $\beta$-sites, such as in light-lanthanide palladides. Probes that start on R-sites, with nominally zero EFG, can only make near-neighbor jumps to Pd-sites, followed by subsequent jumps to either type of sites. The changes in EFGs will produce nuclear relaxation, with the first jump from a site with zero EFG to one with the axially symmetric EFG for probes on $\alpha$-sites. This has not heretofore been theoretically modeled to our knowledge. Such relaxation might already have been observed; for example, consider the spectrum measured at 673 K in Fig. 7. In addition to the signal with 55-ns period for probes on $\alpha$-sites, there is a time-independent offset that can be attributed to probes on cubic sites as well as component with an exponential decay of about 200 ns. The latter signal may arise from probe atoms on cubic R-sites that transfer to the noncubic Pd-sites and then make additional jumps. Since the spectrum at 473 K exhibits no inhomogeneous broadening, the "exponential decay" line-shape contribution at 673 K is most likely caused by diffusional relaxation of the quadrupole interaction of probes on the (cubic) R-sublattice. Relaxation of quadrupole interactions for probes starting on cubic sites may be a useful subject for future experimental and computational study.

**Summary.** Studies were carried out on series of rare-earth indides and palladides having the $L1_2$ structure in order to measure diffusional nuclear relaxation of the quadrupole interaction at nuclei of PAC probe atoms and to determine jump-frequencies and their systematics. Two distinct relaxational regimes were observed along the series of indides. Measurements along the series of palladides revealed strong changes in site-preferences of parent In-probes and similar changes in site-preferences of the daughter Cd-probes were suggested by the relaxation rates. After asserting that a similar change in site-preferences of Cd-probes exists along the indide series, a hypothesis was developed to explain the relaxational regime observed for indides of La, Ce, Pr and Nd: that Cd-probes originally on In-sites transfer to rare-earth sites via an interstitial mechanism. The hypothesis explains the observed composition dependence of the nuclear relaxation in the light-lanthanide indides and also the very large magnitude of the relaxation in $In_3La$ and other phases.

**Acknowledgments.** I wish to thank all who participated in these studies, including F. Selim, A. Favrot, R. Newhouse, J. Bevington, L. Kang, E.R. Nieuwenhuis, D. Solodovnikov, J. Wang, M. Lockwood Harberts, B. Norman, X. Jiang, Q. Wang, and, especially, my long-standing collaborator M.O. Zacate. This work was supported in part by the National Science Foundation under grant DMR 14-10159 and predecessor grants (Metals Program).